\pdfoutput=1
\documentclass[11pt]{article}

\usepackage[table]{xcolor}
\usepackage{acl}

\usepackage{times}
\usepackage{latexsym}
\usepackage{graphicx}
\usepackage{booktabs} % for \specialrule command

\newcolumntype{L}[1]{>{\raggedright\arraybackslash}p{#1}}

\usepackage[T1]{fontenc}
\usepackage[utf8]{inputenc}

\usepackage{microtype}

\usepackage{inconsolata}

\usepackage{fontawesome}

\newcommand{\assistant}{{\texttt{Assistant}}}

\title{Evaluation and Continual Improvement for an Enterprise AI Assistant}

\author{Akash V. Maharaj, 
Kun Qian, 
Uttaran Bhattacharya, 
Sally Fang, 
{\bf Horia Galatanu,} \\
{\bf Manas Garg,} 
{\bf Rachel Hanessian,} 
{\bf Nishant Kapoor,}
{\bf Ken Russell,} \\
{\bf Shivakumar Vaithyanathan, and}
{\bf Yunyao Li} \\
  Adobe Inc. \\
{\tt \{maharaj, kunq, ubhattac, xinf, horiag, mangarg, hanessia\}@adobe.com}\\
{\tt \{niskapoor, kenrusse, vaithyan, yunyaol\}@adobe.com}\\
  }

\begin{document}
{\makeatletter\acl@finalcopytrue
  \maketitle
}
\begin{abstract}
The development of conversational AI assistants is an iterative process with multiple components. As such, the evaluation and continual improvement of these assistants is a complex and multifaceted problem. This paper introduces the challenges in evaluating and improving a generative AI assistant for enterprises, which is under active development, and how we address these challenges. We also share preliminary results and discuss lessons learned. 
\end{abstract}

\section{Introduction}
\label{sec:intro}
%\subsection{AI Assistant high-level overview}

\begin{figure*}[t]
    \centering
    \includegraphics[width=0.8\textwidth]{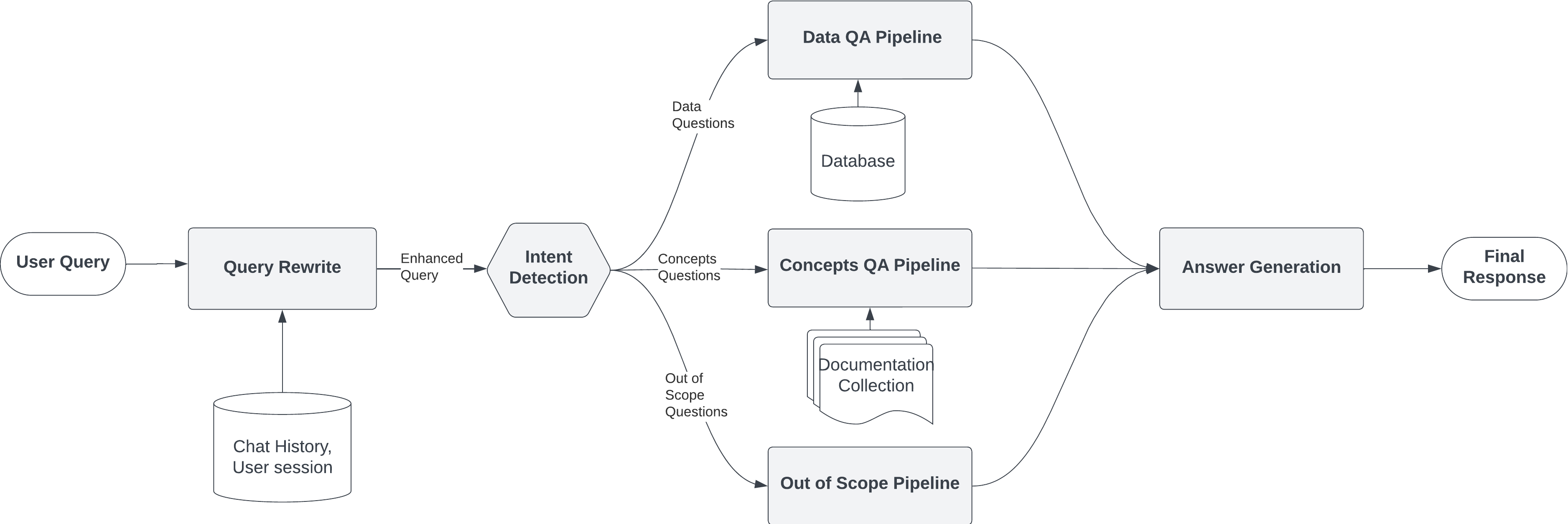}
    \caption{{\assistant} Overall Architecture}
    \label{fig:architecture}
\end{figure*}

Generative AI assistants for enterprises hold the great promise of significantly improved productivity, lowered barrier-to-entry, drastically increased product adoption, transformative amplification of creativity, and delivery of better customer and employee experiences~\cite{generative_bot_ieee_2023}. Developing such an AI assistant is typically an iterative process, with its evaluation and continual improvement at the center. 

Fig.~\ref{fig:architecture} depicts the high-level architecture of Adobe Experience Platform AI Assistant\footnote{Hereafter referred to as \assistant} \cite{bhambhri_2024}, a generative AI assistant built for an enterprise data platform. As can be seen, it is a complex pipeline with multiple underlying components consisting of one or more machine learning models based on large language models (LLMs) or small language models (SLMs). Users interact with the system via a conversational interface to obtain answers based on heterogeneous data sources.
The evaluation and continual improvement of such a system is a complex and multifaceted problem with the following key challenges. 

\noindent\textbf{Metrics.} The success of {\assistant} is ultimately measured by metrics such as user engagement, user satisfaction, and user retention. However, such metrics are lag measures obtainable only after building and deploying {\assistant} in production. To guide continual improvement of {\assistant}, we also need to define metrics that are \textit{lead} measures for various aspects of {\assistant} that are likely to impact the lag measures.

\noindent\textbf{Data.} To produce reliable evaluation metrics for {\assistant}, we need data that are both representative and high-quality. We need a systematic approach to obtain such high-quality data at scale. 

\noindent\textbf{Dynamics.} As shown in Fig.~\ref{fig:architecture}, a real-world AI assistant usually consists of a complex pipeline of components. Each component evolves over time as both the underlying models and the assistant's functionalities change. Further, in enterprise settings, the distribution of questions asked is ever-changing as the customer base shifts and grows and existing customers mature in their adoption of the assistant. We need to consider such customer dynamics.

\noindent\textbf{Human-Centered Design.} The success of {\assistant} depends on both the capabilities of its underlying components and the user interface (UI) that surfaces those capabilities to support the overall user experience. As such, the evaluation and continual improvement for {\assistant} need to take all underlying components as well as UI into consideration for such a human-centered system~\cite{liao2023ai}.  

%\vspace{0.05in}
\noindent\textbf{Privacy and Security.} Enterprise AI assistants like {\assistant} often deal with sensitive user data. We need to evaluate its performance while securely handling customer data and prevent unauthorized access or misuse~\cite{WU2023,YAO2024100211}.

The rest of this paper presents our proposed solution for addressing these changes. We also share our preliminary results and discuss lessons learned so far. Our main contributions include:

\begin{itemize}
    \item A comprehensive continual improvement framework to support the evaluation and continual improvement for {\assistant}. 
    \item A taxonomy of error types for error analysis and continual improvement.
    \item Identifying the limitations of existing approaches on the evaluation of AI assistants.
    \item Highlighting the influential role of human-centered UI design in the evaluation and continual improvement of {\assistant}.
     \item Productionizing such a framework, sharing initial results and lessons learned.
\end{itemize}

% \vspace{-2mm}
\section{Limitations of Existing Approaches}
\label{sec:limitations}
% \vspace{-2mm}
%The evaluation of AI assistants is an essential step in building any human-facing AI system. It serves as the cornerstone for ensuring that these systems meet the standards of reliability and fairness required for human interactions. As evidenced by LLM-based AI systems, which tend to hallucinate inaccurate information and may generate harmful and misleading responses when given adversarial prompts~\cite{zhu2023promptbench}, the importance of evaluating AI systems become apparent. 

% Seeking explicit and implicit feedback from human have been proven to be effective for evaluating and improving intelligent natural language generation systems~\cite{feedback_tacl}. Implicit feedback is gathered passively from user actions within a system, such as clicks, views, or navigation patterns, providing insights into user behavior and preferences without requiring direct input. In contrast, explicit feedback is actively obtained through direct prompts or questions presented to users, allowing them to express their opinions, preferences, and experiences explicitly. 

% Another crucial aspect of evaluating AI assistants is benchmarking. Utilizing benchmark datasets, particularly domain-specific ones, is a typical way to assess the performance of AI assistnats in the enterprise scenarios where domain-specific knowledge needs to be handled. 

Common approaches for evaluating AI assistants include evaluation using explicit feedback, evaluation using implicit feedback, benchmarking~\cite{liang2022holistic}, and human evaluation~\cite{feedback_tacl}. Explicit feedback is collected from users through feedback buttons, direct prompts, or questions on their preferences. In contrast, implicit feedback is gathered from user actions within a system, such as clicks, views, or navigation patterns, providing insights into user behavior and preferences without requiring direct input. Evaluating with benchmark datasets is also a common way to evaluate AI assistants.
% proves to be an effective method for assessing the performance of AI assistants, especially in enterprise scenarios where they need to handle domain-specific knowledge.
%As we will discuss shortly, all these methods face challenges and limitations when it comes to enterprise AI assistant scenarios.
These approaches, while important and effective to a certain degree, suffer from various limitations when it comes to evaluating an enterprise AI assistant such as {\assistant}, which is under active development and improvement.

\subsection{Limitations of Explicit Feedback}
%Collecting explicit user feedback is an effective way to gauge user satisfaction and improve the performance of an AI assistant, which may sound straightforward and simple, but it faces challenges. Summarized from an earlier iteration of our user study, Table \ref{table:explicit_feedback} illustrates the feedback distribution from customers who interacted with our AI assistant. Several limitations and challenges can be observed: (1) \textit{low engagement rate} (2) \textit{skewed feedback distribution} , and (3) \textit{lack of detailed feedback}. 
% Collecting explicit user feedback seems to be the most straightforward way to gauge user satisfaction and gather input to improve the performance of an AI assistant. However, we observe three limitations of this approach based on the initial set of explicit feedback for {\assistant} (Table \ref{table:explicit_feedback}): (1) \textit{sparsity} (2) \textit{skewed feedback distribution} , and (3) \textit{lack of detailed feedback}. 
Collecting explicit feedback from the users seems to be the most straightforward way to gauge user satisfaction and gather input to measure and improve the performance of an AI assistant. Table \ref{table:explicit_feedback} illustrates the initial set of explicit feedback for {\assistant} from our early customers. We can observe several limitations of this approach. 
%(1) \textit{sparsity} (2) \textit{skewed feedback distribution} , and (3) \textit{lack of detailed feedback}.

% \textbf{Low engagement rate}: Table \ref{table:explicit_feedback} reveals we did not obtain any feedback for 76\% of all customer interactions. This low engagement ratio makes it challenging for us to understand user experience and satisfaction comprehensively and hampers efforts to improve the AI assistant's performance.
\noindent\textbf{Sparsity.} Explicit user feedback is sparse. From Table \ref{table:explicit_feedback}, we can see that 76\% of all customer interactions receive no explicit feedback at all. This sparsity issue makes it challenging to understand user experience and satisfaction comprehensively and hampers efforts to improve {\assistant}.

% \textbf{Skewed feedback distribution}: Since sharing feedback is not mandatory, not every customer provided their feedback to us. Customers with more interactions or those who are willing to share may be more inclined to provide feedback. As shown in Table \ref{table:explicit_feedback}, there are two customers who did not share any feedback with us. On the other hand, although not explicitly illustrated in Table \ref{table:explicit_feedback}, Customer {\tt Org1} contributed about 30\% of all the feedback we collected from all customers. Obviously, the skewed feedback distribution can potentially misrepresent the overall sentiment towards the AI assistant, leading to an inaccurate reflection of the diverse range of users' experiences and opinions.

\noindent\textbf{Representativeness.} Since sharing explicit feedback is not mandatory, not every user does so. As shown in Table \ref{table:explicit_feedback}, users from two organizations shared no feedback at all. Further examination showed that most feedback came from a small number of users. In fact, about 30\% of all the feedback originated from one user. Such a highly skewed feedback distribution may misrepresent the overall sentiment towards \assistant, and fail to reflect the diversity of users' experiences and opinions.

\noindent\textbf{Lack of detailed feedback.} partly due to minimizing user effort and partly because users only see the final response, explicit feedback is usually gathered via a simple UI form (\textit{e.g.}, \faThumbsOUp, {\faThumbsODown} buttons). Unfortunately, feedback gathered this way often fails to capture the nuances of user experiences and preferences. For instance, a negative feedback indicating an incorrect final response is insufficient to pinpoint specific components for improvement. New approaches like showing step-by-step explanations and getting user feedback for the explanation are alternative ways to get detailed feedback from users and map them to specific components.

\begin{table}[t]
\fontfamily{cmss}\selectfont
\small
\centering
\caption{Feedback type distribution and engagement ratio from different customers}
%\vspace{-1mm}
\label{table:explicit_feedback}
%\begin{small}
\begin{tabular}{lrrr}
\toprule
\multicolumn{1}{c}{\textbf{Customer}} & \multicolumn{1}{c}{\textbf{\begin{tabular}[c]{@{}c@{}}Positive \\ feedback\end{tabular}}} & \multicolumn{1}{c}{\textbf{\begin{tabular}[c]{@{}c@{}}Negative \\ feedback\end{tabular}}} & \multicolumn{1}{c}{\textbf{\begin{tabular}[c]{@{}c@{}}No \\ feedback\end{tabular}}} \\
\midrule
Org1                                  & 22.8\%                                                                                    & 16.2\%                                                                                    & 61\%                                                                                \\
Org2                                  & 12.6\%                                                                                    & 11.2\%                                                                                    & 76.2\%                                                                              \\
Org3                                  & 3.2\%                                                                                     & 24.9\%                                                                                    & 71.9\%                                                                              \\
Org4                                  & 2.7\%                                                                                     & 5.0\%                                                                                     & 92.3\%                                                                              \\
Org5                                  & 11.3\%                                                                                    & 5.2\%                                                                                     & 83.5\%                                                                              \\
Org6                                  & 5.6\%                                                                                     & 9.7\%                                                                                     & 84.7\%                                                                              \\
Org7                                  & 8.6\%                                                                                     & 21.4\%                                                                                    & 70\%                                                                                \\
Org8                                  & 15.4\%                                                                                    & 7.7\%                                                                                     & 76.9\%                                                                              \\
Org9                                  & 0\%                                                                                       & 0\%                                                                                       & 100\%                                                                               \\
Org10                                 & 0\%                                                                                       & 0\%                                                                                       & 100\%                                                                               \\
\midrule
\textbf{Total}                        & \textbf{10.72\%}                                                                          & \textbf{13.12\%}                                                                          & \textbf{76.16\%}                                                                    \\ 
\bottomrule
\end{tabular}
%\end{small}
%\vspace{-5mm}
\end{table}

\subsection{Limitations of Implicit Feedback}
% Implicit feedback is an important way to compensate the lack of explicit feedback, which has been extensively used in evaluating and improving intelligent systems (\textit{e.g.},~\cite{implicit_for_recommender_system, koren2021advances}). Since implicit feedback is obtained indirectly and passively from customer actions, it may not always reflect users' true preferences. As a result, prior work usually use denoising techniques to  prune the noisy interactions to avoid serious negative impact~\cite{wang2021denoising}.  Moreover, deriving implicit feedback from user interactions may pose challenges depending on the tasks performed by AI assistants. For instance, while implicit feedback is readily available for recommender systems in contexts such as online shopping, where users engage with various aspects of the system (such as browsing history, mouse clicks, page view duration, scrolling behavior, and add-to-cart actions), obtaining meaningful implicit feedback for AI assistants primarily handling text-based user queries may be more difficult. Verifying the occurrence of a rephrased follow-up question is a valuable form of implicit feedback. However, obtaining other indicative signals is challenging due to the inherently simplistic nature of the required user interactions. Even if we are able to obtain implicit feedback with positive or negative sentiments, we still face the challenge of lacking detailed feedback.
Implicit feedback has been extensively used in evaluating and improving intelligent systems (\textit{e.g.},~\citet{implicit_for_recommender_system, koren2021advances}), and performance measurements of concrete tasks have been recommended as the best metric for evaluating natural language generation systems~\cite{saphra2023first}. This approach has several limitations when evaluating AI assistants. First, since implicit feedback is obtained indirectly and passively from user actions, it may not always reflect users' true preferences. Prior work uses denoising techniques to prune the noisy interactions to avoid serious negative impact~\cite{wang2021denoising}. In addition, deriving implicit feedback from user interactions could be a challenge on its own. For instance, while meaningful implicit feedback is readily available for recommender systems in contexts such as online shopping (clicks, page views, add-to-cart, etc.), implicit signals available in AI assistants are less clearly related to concrete user goals. Specifically, users have a wide variety of goals, and the concrete tasks to achieve those goals are often very delayed. 

\subsection{Limitations of Off-the-Shelf Benchmarks}
Although public benchmark datasets for general tasks are abundant (\textit{e.g.},~\citet{chang2023survey} lists 46 public benchmark datasets), they are often not applicable for domain-specific AI assistants. Creating domain-specific benchmark datasets is labor-intensive, time-consuming, and requires domain expertise. Moreover, assistants' workload and tasks may also evolve. Thus, there is no one static benchmark data that suits all~\cite{mizrahi2024state}. Therefore, benchmark data creation itself is a continual process.

\begin{table*}[t]
    \fontfamily{cmss}\selectfont
    \centering
    \caption{Error Severity Framework in \assistant}
    %\vspace{-2mm}
    \label{tab:errorseverity}
    \resizebox{\textwidth}{!}{%
    \begin{tabular}{L{0.15\hsize}L{0.2\hsize}L{0.2\hsize}L{0.45\hsize}}
        \toprule
        Category & Definition & Consequence & Examples\\
        \toprule

        \textbf{\textcolor{red}{Severity 0}} & \small{\textbf{Answer looks right, but is wrong}}&
        \small{\textit{\textbf{Erodes trust with the users}}}&
        \begin{minipage}[t]{0.4\textwidth}%
        \small{
        \noindent- Convincing Concepts QA answers that are pure hallucinations\\
        %\vspace{0.5pt}
        \noindent- Incorrect Data QA answers that cannot easily be verified independently\\
        }
        \end{minipage} \\

        \arrayrulecolor{gray}
        \midrule
        
        \textbf{\textcolor{magenta}{Severity 1}} & \small{\textbf{Answer looks wrong, user can't recover}}&
        \small{\textit{\textbf{Frustrates users}}} &
        \begin{minipage}[t]{0.4\textwidth}%
        \small{
        \noindent- Failure to answer with generic error message \\
        %\vspace{0.5pt}
        \noindent- Answers with obvious logical inconsistencies, \textit{e.g.}, mixing UI docs and API docs
        }\\
        \end{minipage} \\

        \midrule
        
        \textbf{\textcolor{orange}{Severity 2}} & \small{\textbf{Answer looks wrong, user can recover}} &
        \small{\textit{\textbf{Annoys users}}} &
        \begin{minipage}[t]{0.4\textwidth}%
        \small{
        \noindent- Misunderstood questions that user is able to rephrase and get correct answer \\
        \noindent- Incorrect out-of-scope question rejection that user is able to override
        }
        \end{minipage} \\
        \arrayrulecolor{black}
        \bottomrule
    \end{tabular}
    }
\end{table*}

% \vspace{-10mm}
\section{Our Approach}
\label{sec:ourapproach}
% \vspace{-2mm}
In this section, we introduce our framework to overcome the aforementioned challenges (Section~\ref{sec:intro}) and limitations of existing approaches (Section~\ref{sec:limitations}) for evaluating an enterprise-grade AI assistant under active development.

\subsection{Design Decisions}
%When designing any such evaluation framework in a fast moving product there always trade-offs to be made, both in terms of breadth and depth of any given type of evaluation. It is therefore crucial to define a concrete set of design goals that can help in making such trade-offs. We have focused on the following goals:
We first present a few key design decisions to balance the trade-offs to be made, both in terms of breadth and depth of any given type of evaluation. 

%\vspace{0.05in}
\noindent\textbf{Prioritize metrics directly impacted by production changes.} The ultimate goal of {\assistant} is to improve the productivity and creativity of our users and lower barriers to entry. Since it takes time to materialize such lag measures, we focus on directly responsive ``correctness'' metrics, assuming a more correct {\assistant} will ultimately lead to positive downstream outcomes. 

%\vspace{0.05in}
\noindent\textbf{Align metrics with user experience.} Not all errors are equal. The impact on the user experience of one incorrect citation in an otherwise correct answer is very different from that of a completely hallucinated answer. We aim to capture this nuance in the design of our error metrics.

%\vspace{0.05in}
\noindent\textbf{Human Evaluation over automated evaluation.} We believe that, despite challenges~\cite{clark-etal-2021-thats}, human judgments are still best aligned with eventual user outcomes. As such, we prioritize human evaluation over automated evaluation. Once high-quality human judgments are collected, they can be used to validate which automatic evaluations are meaningful for specific tasks and components.

%\vspace{0.05in}
\noindent\textbf{Efficient allocation of human evaluators.} To conduct human evaluation at scale, we focus on the efficient allocation of human annotators. Specifically, simple annotation tasks are done by non-experts, while complex error analysis and the determination of how to make improvements are left up to engineers with domain expertise. 

%\vspace{0.05in}
\noindent\textbf{Collect both end-to-end metrics and component-wise metrics.} We collect both individual and collective metrics to understand the overall quality of the system as well as which parts need to improve.

%\vspace{0.05in}
\noindent\textbf{System-wide improvements.} All components in {\assistant}, from ML/rule-based models, UI/UX components, to underlying data, may impact system performance. Therefore, instead of focusing solely on ML model improvements, we consider the entire ``vertical" system holistically and leave no improvement off the table. 

%\vspace{0.05in}
\noindent\textbf{Prioritize human evaluation.} Automated evaluation, which utilizes standard metrics and evaluation tools, is popular for its efficiency and objectivity~\cite{chang2023survey}. 
However, although more labor-intensive and time-consuming, manual evaluation by domain experts is more reliable in reflecting the final user impact. As such, we prioritize human evaluation over automation.

\subsection{Severity-based Error Taxonomy}

Designing metrics that align with our end users' judgments of the correctness and usefulness of {\assistant} is a complex task. We observed relatively high error rates from an early version of \assistant~(over 50\%), yet our users did not seem to perceive error rates to be this high in their self-reported surveys and regular feedback sessions. This discrepancy, consistent with the earlier observation that not all errors are the same~\cite{freitag2021experts}, led us to develop a \textit{taxonomy} of errors.   

To illustrate this point, consider the past two decades, where internet search has become a dominant (semi) natural language interface. In this domain, humans have become accustomed to certain classes of errors. When we do not get the desired results from a search engine, we rephrase and iterate till we find the answer. The initial failure of the search engine is annoying but generally tolerable unless we cannot find our answer even after many re-phrasings. At this point, we are left frustrated. Inspired by DevOps terminology~\cite{kim2021devops}, we can define two separate classes of errors: \textbf{Severity-2} (``Sev-2'' for short) errors are annoying but repairable via rephrasing, while \textbf{Severity-1} (``Sev-1'' for short) errors are not repairable.

Meanwhile, the rise of generative AI has introduced an entirely new class of error: answers that are convincing and look correct but are, in fact, wrong. Depending on the use case, these may be tolerable (or even desirable), but in the realm of enterprise assistants, these errors are troubling. They erode user trust and may lead to complete abandonment of the assistant. We term these \textbf{Severity-0} errors, ``Sev-0'' for short. Table~\ref{tab:errorseverity} summarizes this severity-based error taxonomy, which has become an organizing principle for the evaluation and improvement of \assistant, as we discuss next.

\subsection{Framework for Evaluation and Continual Improvement}
\begin{figure*}[t]
    \centering
    \includegraphics[width=.85\textwidth]{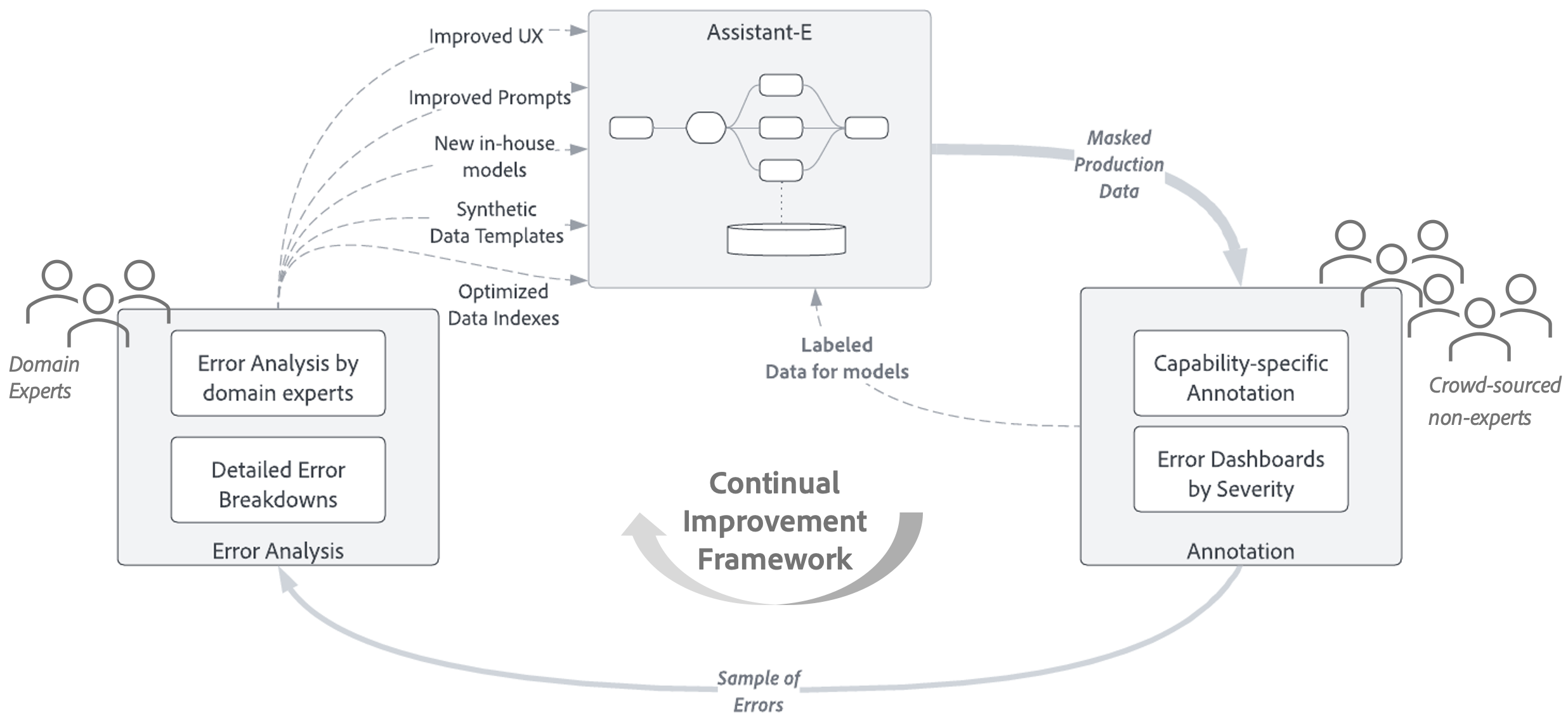}
    \caption{Evaluation and continual improvement framework of {\assistant}}
    \label{fig:framework}
\end{figure*}

Fig.~\ref{fig:framework} depicts our proposed evaluation and improvement framework. It includes three main components: \assistant, itself, a dedicated Annotation Tool, and a separate environment for Error Analysis. Human evaluation drives the evaluation and improvement of \assistant. 

To ensure the efficient allocation of human resources, non-experts provide large-scale annotation of masked production data, while domain experts provide detailed error analysis on a sample of production data. For each annotation task, to ensure annotation quality, we design the UI and annotation guidelines iteratively with pilot study and improvements. We include training modules and exercises to ensure annotators meet a minimum bar of sufficient domain understanding. We assign multiple annotators for each annotation task to further ensure the annotation quality and conform to best practices~\cite{van2021human}.

We design different annotation tasks to assess the quality of different \assistant~components and improvements needed. By collecting annotations based on prior interactions in the production system, we can generate both error metrics by severity (by comparing human labels to the choices the system made in production), \textit{and} new golden-labeled data for model improvements.

%While the primary outputs of annotation are the component wise error metrics and new labeled data for training in-house models, the 

Error analysis is a crucial step in gating improvement. At this step, domain experts --- those with deep knowledge of how \assistant~is designed --- review samples of errors, identify error patterns, and determine specific improvements. These improvements take many potential forms, from prompt engineering to training and improving in-house models, to creating new templates and patterns for synthetic data, to more holistic changes such as improving the user experience or optimizing the specialized data indexes that are queried by \assistant, (for example, fine-tuning embeddings, or updating database schema). This last category of improvements is only possible when the application is viewed holistically, and all stakeholders are involved in error analysis. 

%\vspace{-2mm}
\section{Preliminary Results: Examples}
\label{sec:results}
%\vspace{-2mm}
While \assistant~remains in active development, our evaluation and csontinual improvement framework already show promising impacts on both the prioritization and the design of improvements. In this section, we share the preliminary results obtained so far by examples.

\begin{figure}[!b]
    \centering
    \includegraphics[width=0.9\columnwidth]{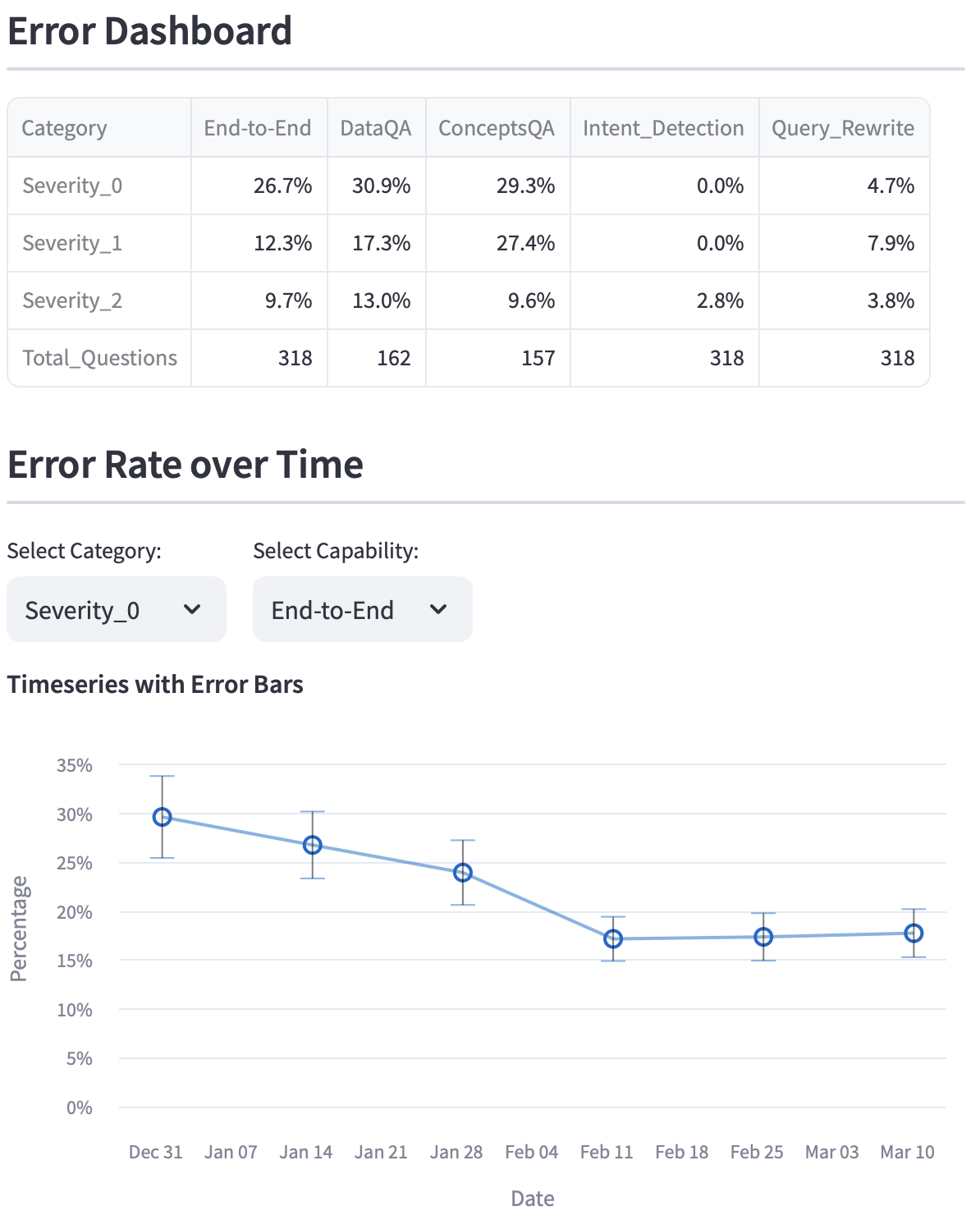}
    \caption{Dashboard showing snapshot of Error Severities and time-evolution for a single component. Illustrative data of similar magnitude to production numbers.}
    \label{fig:errordashboard}
\end{figure}

Fig.~\ref{fig:errordashboard} illustrates an example error dashboard produced by the annotation tool, showing component-wise and end-to-end errors with further breakdown by severity levels as well as how they change over time. This dashboard is monitored by all stakeholders and is used to track the impact of feature releases and improvements. 
While ideas for improving \assistant~may be endless, detailed error analysis allows the team to follow a powerful organizing principle: focusing on reducing error rates based on their actual impact on the users.

% For instance, the example report in Fig.~\ref{fig:erroranalysis} highlights hallucination and document retrieval as the major areas for improvement for Concept QA component, after an Out-of-Scope detector was added for Concept QA questions to reduce out-of-scope errors as dominant Sev-0 errors. 

% We can also see that \textit{OutOfScope}-\textit{Question-Incorrectly-Rejected} appears as a Sev-2 errors reflects a major success of this evaluation framework.  However, a major concern was to not incorrectly reject in-scope questions. By thinking holistically about the application, we worked with our UX partners to design an override mechanism for users that would allow them to receive an answer. This converted a potential Sev-1 error (refusal to answer), into a Sev-2 error (the user could now recover). This is a great example of how human centered UI design allows holistic improvement of \assistant. 
For instance, the example report in Table~\ref{table:error_summary} shows that Out-of-Scope errors were our largest contributor to Sev-0 errors in Sprint 1. To address this, we introduced an Out-of-Scope text classifier using an in-house model, which achieved 90\% precision and successfully reduced most such errors. 

However, the new classifier also led to a new, particularly frustrating source of errors: in-scope questions misclassified as out-of-scope would no longer be answered. Without being able to quickly improve the classifier's precision, we used our other available lever of improvement and designed an override mechanism in the UI to allow users to receive an answer. As the Sprint 2 report shows, this UI change converted a potential Sev-1 error (refusal to answer) into a Sev-2 error (the user could now recover), showcasing how human-centered UI design allows holistic improvement of \assistant.

\begin{table}[t]
\fontfamily{cmss}\selectfont
\centering
\begin{small}
\caption{An example output of error analysis for Concept QA (illustrative data, real labels) from one sprint to the next after Out-Of-Scope detection was deployed.}
\label{table:error_summary}
\begin{tabular}{lrr}
\toprule
\textbf{Error Severity} & \textbf{\% Sprint1} & \textbf{\% Sprint2} \\
\toprule
\textbf{Sev-0} & 53.4\% & 36.6\% \\
\quad \textit{OutOfScope} & \textcolor{red}{\textbf{21.6}\%} & \textcolor{red}{\textbf{6.2}\%}\\
\quad \textit{Hallucination} & 17.0\% & 16.4\%\\
\quad \textit{Doc-Retrieval} & 13.6\% & 14.0\%\\
\quad \textit{LLM-Error} & 1.1\% & 0.0\% \\
\midrule
\textbf{Sev-1} & 46.6\%  & 44.4\% \\
\quad \textit{Hallucination} & 36.4\% & 33.0\% \\
\quad \textit{Citation} & 5.7\% & 5.1\% \\
\quad \textit{LLM-Error} & 4.5\% & 6.3\% \\
\midrule
\textbf{Sev-2} & - & 6.9\%\\
\quad \textit{OutOfScope} & - & \textcolor{blue}{6.9\%}\\
\quad\textit{(incorrect rejection)}&&\\
\bottomrule
\end{tabular}
\end{small}
\end{table}

% Another attempt we made to reduce the severity of Sev-0 errors is to explain the results we present to our customers. Explainability is an important technique for improving user trust and comprehension. However, there are many explainability techniques~\cite{survey2020} that can be adopted to generate various types of explanations. Choosing which ones to use given the time and resource constraints we face is crucial. In practice, we took a data-driven approach, where we first identified a list of eight different potential techniques. Then, domain experts went through the Sev-0 queries obtained during a certain window to manually annotate which technique can be used to alleviate the severity of the Sev-0 query. Finally, the potential impact of each explainability technique can be visually illustrated through a histogram (\textit{e.g.}, Fig. \ref{fig:exp_histogram}). In addition to the potential impact, we also need to consider the implementation difficulty and human cognitive load to make a holistic decision that would give us the best return.

Explainability is important for improving user trust and comprehension. By helping users discover wrong answers with better explainability, we can reduce Sev-0 errors and move them to Sev-1/Sev-2 error buckets. We took a data-driven approach to choose from many applicable explainability techniques~\cite{survey2020}. We first went through the Sev-0 queries obtained during a certain window and examined which technique(s) can be used to alleviate the severity of each error based on the potential overall impact of each explainability technique, its implementation difficulty, and human cognitive load. We created a decision matrix (Table~\ref{table:decision_matrix}) based on the analysis, and we focused on only 2 of the 7 options from~\cite{nlidb-book}. As we move forward, we expect many more such informed improvements based on our framework.

%As we move forward, we expect many more examples of this kind of informed improvement to result from our framework. 

% \begin{itemize}
%     \item{What does dashboard look like - rough}
%     \item{Examples - how it changes over time- two or three iterations}
%     \item{Error analysis breakdown of documentation errors, and how this prioritizes improvement}
% \end{itemize}

%\vspace{-2mm}
\section{Discussion}

This framework has organically evolved during the development of \assistant. While many of the design choices laid out may seem obvious in hindsight, they were not as clear at the beginning of this project, and so it is worth discussing the lessons we have learned along the way.

First, we have found that metric design is of paramount importance. The severity framework came after many iterations in trying to connect enthusiastic early customer feedback with a seemingly large overall error rate. The insight that customers have varying tolerance depending on the class of errors has become a powerful organizing principle for our prioritization and resource allocation to improve \assistant. 

Next, we have seen firsthand the benefits of building a decomposed system as opposed to depending on a single, monolithic model. The choice to decompose into multiple, orchestrating models was led by constraints such as task specialization and the need to query real-time data. We have also reaped the secondary benefit of having many available ``levers of improvement'' (prompts, in-house models, specialized indexes, UX improvements, etc.), many more than what is possible in a single language model paradigm.

% \looseness=-1 With an AI assistant composed by multiple models chaining together, we faced the challenge of creating both end-2-end and component-wise benchmarks for proper evaluation. It follows that, conducting evaluations in a continual fashion requires considerable effort, as even minor changes to components can drastically alter the final responses. Therefore, a robust evaluation framework must be established to ensure the accurate and comprehensive assessment of AI assistants' performance over time.

Finally, iterative and agile development are more important than designing everything upfront and building specialized tools. For instance, while it is tempting to build in-house tools, using spreadsheets as a simple alternative initially allows us to learn important lessons on designing the annotation tasks, from annotation guidelines to the actual UI. 

\begin{table}[t]
\fontfamily{cmss}\selectfont
\small
\caption{Decision matrix for explainability techniques}
%\vspace{-2mm}
\label{table:decision_matrix}
\begin{tabular}{lccc}
\toprule
\multicolumn{1}{c}{\textbf{\begin{tabular}[c]{@{}c@{}}Explainability\\ techniques\end{tabular}}} & \multicolumn{1}{c}{\textbf{\begin{tabular}[c]{@{}c@{}}Potential \\ impact\end{tabular}}} & \multicolumn{1}{c}{\textbf{\begin{tabular}[c]{@{}c@{}}Engineering\\ difficulty\end{tabular}}} & \multicolumn{1}{c}{\textbf{\begin{tabular}[c]{@{}c@{}}Congitive\\ load\end{tabular}}} \\
\toprule
technique1                                                                   & 0.0\%                                                                                    & high                                                                                          & low                                                                                   \\
technique2                                                                   & 8.6\%                                                                                    & high                                                                                          & high                                                                                  \\
\rowcolor[HTML]{9AFF99} 
technique3                                                                   & 48.6\%                                                                                   & low                                                                                           & low                                                                                   \\
\rowcolor[HTML]{9AFF99} 
technique4                                                                   & 88.6\%                                                                                   & medium                                                                                        & medium                                                                                \\
technique5                                                                   & 20.0\%                                                                                   & high                                                                                          & low                                                                                   \\
technique6                                                                   & 100\%                                                                                    & medium                                                                                        & low                                                                                   \\
technique7                                                                   & 74.3\%                                                                                   & high                                                                                          & low \\                \bottomrule                
\end{tabular}
\end{table}

% \vspace{-2mm}
\section{Future Work}
% \vspace{-2mm}
As we continue to develop {\assistant} and onboard more customers, we plan to extend our evaluation and continual improvement framework with more human-in-the-loop/LLM-in-the-loop automation to scale our evaluation and error analysis processes~\cite{NEURIPS2023_91f18a12}. In addition, the current framework heavily focuses on retrospective analysis based on \textit{past} customer interactions. We plan to extend it with more proactive user studies and evaluation of in-development functionalities. Moreover, personalization is also important for enterprise AI assistants since we have customers with different technical levels. To provide the best experience to various personas in potentially different languages, additional evaluation metrics and datasets proposed in~\cite{DBLP:journals/corr/abs-1709-04734, ahuja2023mega} may also be considered. As we have emphasized, human-centered design is essential for the success of {\assistant}. We plan to further explore how the deeper interplay between ML and UX components in this new paradigm of HCI can lead to more explainable and accurate assistants. Finally, the impact of generative AI applications in the workplace is an important new area of study~\cite{brynjolfsson2023generative}. As we enroll new customers, we intend to run A/B tests~\cite{hussey2007design} that assess the causal impact of \assistant~on the engagement and productivity of customers.  

\section*{Acknowledgments} We would like to express our gratitude to the engineering team that developed our initial version of the annotation platform: Amy Chen, Kratika Gujrati, Jason Hightower, Saikat Kanjilal, Jayna Punatar, and Siddhartha Sahai. We also acknowledge the foundational work of Saurabh Tripathy in creating the process for human-in-the-loop annotation and error analysis, and for ensuring the right data logging to enable the continual improvement process. 

\bibliography{custom}

% \appendix

% \section{Example Appendix}
% \label{sec:appendix}

% This is an appendix.

\end{document}